\documentclass[conference]{IEEEtran}

\usepackage{cite}
\usepackage[pdftex]{graphicx}
	\ifCLASSINFOpdf
	 	\usepackage[pdftex]{graphicx}
  	\else
	\fi
\usepackage[cmex10]{amsmath}
\usepackage{amsthm}
\usepackage{array}
\usepackage{amssymb}

\begin{document}

\title{A Delay-Constrained General Achievable Rate and Certain Capacity Results for UWB Relay Channel}

\author{\IEEEauthorblockN{Maryam Faramarzi Yazd }
\IEEEauthorblockA{Department of Electrical Engineering\\
Ferdowsi University of Mashhad\\
Mashhad, Iran\\
Email: maryam.faramarzi@gmail.com}
\and
\IEEEauthorblockN{Ghosheh Abed Hodtani}
\IEEEauthorblockA{Department of Electrical Engineering\\
Ferdowsi University of Mashhad\\Mashhad, Iran\\
Email: ghodtani@gmail.com}}

\maketitle

\begin{abstract}
In this paper, we derive UWB version of (i) general best achievable rate for the relay channel with decode-and-forward strategy and (ii) max-flow min-cut upper bound, such that the UWB relay channel can be studied considering the obtained lower and upper bounds. Then, we show that by appropriately choosing the noise correlation coefficients, our new upper bound coincides with the lower bound in special cases of degraded and reversely degraded UWB relay channels. Finally, some numerical results are illustrated.\end{abstract}

\IEEEpeerreviewmaketitle

\section{Introduction}
In a UWB relay channel, two diversity techniques, i.e., cooperative communication and UWB radios are combined and the system performance is improved considerably \cite{Sendonaris2003}.

The relay channel was first introduced in  \cite{vandermeulen-aap-1971}. In \cite{covergamal1979}, the relay channel was studied carefully, e.g., special capacity results and the best achievable rates via two coding schemes, namely decode-and-forward (DF) and estimate-and-forward (EF) strategies were obtained. We have other studies on the relay channel in\cite{gamalaref1982, gamalzahedi2005, aleksicrazaghi2009, cover2007, hodtani2008, hodtani2009}.

Limited research has been done on the capacity bounds for frequency-selective block-fading relay channels. Achievable rates using amplify-and-forward (AF) with network training in which the source node and the destination node broadcast training symbols and each relay node carries out channel estimation are analyzed in  \cite{wang2006}, for narrow and wideband relaying over frequency selective fading channel. In \cite{zolfa2009} an upper bound and DF lower bound were derived for UWB relay channel with an assumption of independent noises at the relay and destination. In \cite{goldsmith2011} Gaussian relay channels with correlated noises have been studied.

\textbf{Our Work}, includes first, the investigation of a more general lower bound that is achieved by partial decode-and-forward scheme (PDF) and determining it for frequency-selective block-fading UWB relay channel. Our result encompasses the DF lower bound in \cite{zolfa2009}. Second, we obtain the UWB new version of max-flow min-cut upper bound with the assumption of correlated noises at the relay and destination. Third, we show that the upper bound coincides with the lower bound in two special cases of degraded and reversely degraded relay channels when the corresponding correlation coefficients are applied and the capacity is determined.

\textbf{Notation:} Throughout the paper $\Re{(.)}$, $\mathbb{E}$, $var(.)$ and $cov(.)$ denote real part, expectation, variance and covariance operations, respectively. $\lfloor x\rfloor$ returns the largest integer $\leq x$. $diag(.)$ builds a diagonal matrix and $C(x)\triangleq\log(1+x)$.

The paper is organized as follows: In Sec. II we define the UWB relay channel model. The lower bound on the capacity of the relay channel obtained via PDF strategy and the max-flow min-cut upper bound for the defined channel model are derived in Sec. III. Two capacity achieving cases corresponding to the degraded and reversely degraded relay channels are discussed in Sec. IV and Sec. V, respectively. Numerical results are illustrated in Sec. VI and finally, we provide the conclusion in Sec. VII.
\setlength{\arraycolsep}{0.0em}
\section{System Model}
\begin{figure}[!t]
\centering
\includegraphics[width=3in]{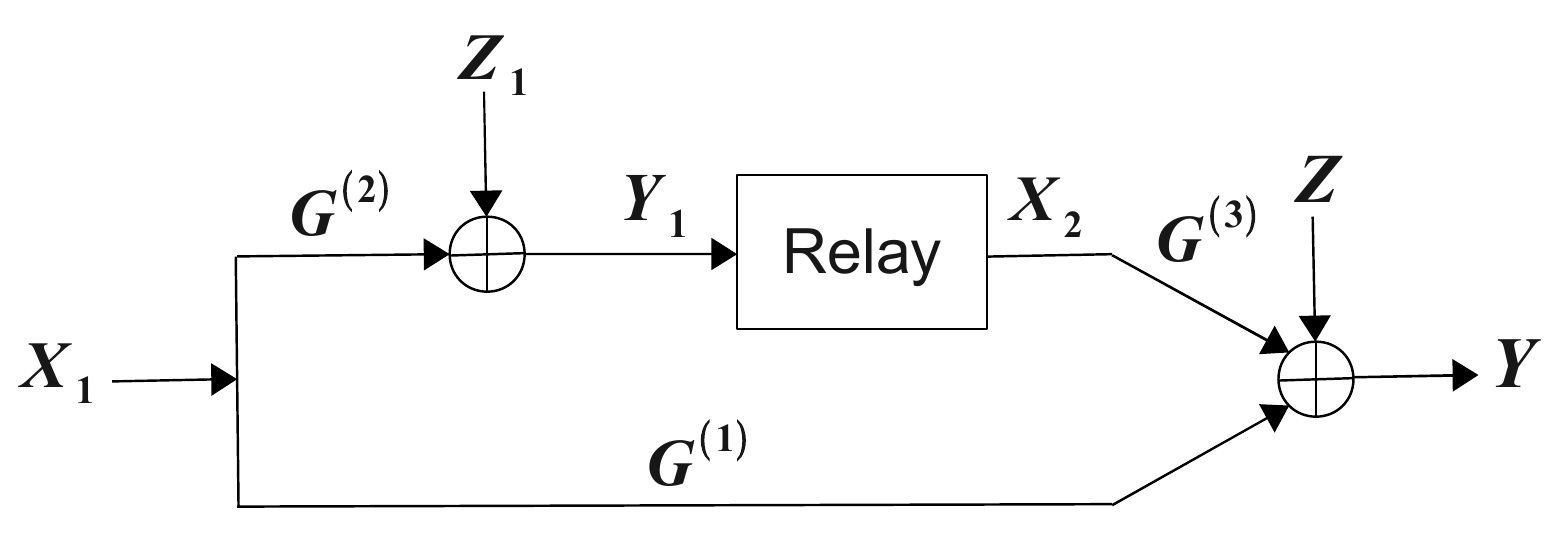}
\caption{Illustration of UWB relay channel.}
\label{figure1}
\end{figure}

We assume that data is sent in blocks of size $K$ as a train of impulse based UWB signals to the destination via link 1 and to the relay via link 2, as depicted in Fig.\ref{figure1}. Based on the received signal, the relay builds a secondary message and forwards it as a UWB signal to the destination via link 3. We assume that all nodes are perfectly synchronized and channel state information (CSI) is available at the receiving terminals only. The fading coefficients between different nodes are assumed mutually independent identically distributed. We assume arbitrarily correlated noises at the relay and destination. The complex baseband impulse response of each UWB link can be considered based on the Saleh-Valenzuela (S-V) model \cite{ieee802.15.4a}
\begin{equation}
h(t){}={}\overset{\sim}{\beta}\sum_{l=0}^{L-1} \sum_{i=0}^{M-1}a_{i,l}e^{j\phi_{i,l}}\delta(t-T_{l}-\tau_{i,l}),
\end{equation}
where $L$ is the number of clusters and $M$ is the number of rays in each cluster. $T_{l}$ and $\tau_{i,l}$ represent the cluster and ray arrival times. The factor $\overset{\sim}{\beta}$ jointly models the pathloss, shadowing, and antenna insertion loss. $a_{i,l}$ is the gain of the $i$th path in the $l$th cluster and finally $\phi_{i,l}$ is the complex baseband phase of each multipath component.\\ 
We know that if the transmitter sends a block of $K$  symbols $(x_{0},\cdots,x_{K-1})^{T}$ through the above UWB channel, the received signal $(y_{0},\cdots,y_{K-1})^{T}$  is in the form \cite{arikan2004}:\\
\begin{equation}
\begin{array}{@{}lc@{\quad}@{\quad}@{\quad}r@{}}
y_{i}=\sum_{k=0}^{K^{'}-1}g_{k}x_{i-k}+z_{i}&&(i=0,\cdots,K-1)
\end{array} \label{eq.2}
\end{equation}
where $z_{i}$s are complex zero mean additive white Gaussian noises, $K^{'}$ is the ISI length due to the multipath fading, and $g_{k}$'s are related to the channel impulse response as\\
\begin{equation}
g_{k}{}={}\sum_{i,l:\lfloor(T_{l}{}+{}\tau_{i,l})/T_{s}\rfloor {}={}k} \overset{\sim}{\beta} a_{i,l}e^{j\phi_{i,l}}.
\end{equation}
In this equation, $T_{s}{}={}\frac{1}{W}$, where $W$ is the bandpass channel bandwidth. We assume that the channel coefficients stays constant within each block of data transmission and change in an independent and identically distributed fashion from one block to another, i.e. a block fading channel is considered because the UWB channel is underspread \cite{ieee802.15.4a}. The size of each block, $K$ is constrained by the channel coherence time and can be at most equal to $\frac{T_{c}}{T_{s}}$. Taking the DFT of the two sides of (\ref{eq.2}), we obtain the frequency domain UWB channel model \cite{arikan2004}. The vectors $\textbf{G}^{(n)}(n=1,2,3)$, $\textbf{X}_{\textbf{1}}$, $\textbf{X}_{\textbf{2}}$, $\textbf{Y}_{\textbf{1}}$ and $\textbf{Y}$ corresponds to the $K$-point DFT of vectors of complex baseband channel coefficients related to each link, the transmitted signals from the source and the relay and the received signals at relay and destination, respectively. Now, we can formulate the input-output relation for the UWB relay channel in the frequency domain as
\setlength{\arraycolsep}{0.0em}
\begin{eqnarray}
Y_{1i}&{}={}&G_{i}^{(2)}X_{1i}+Z_{1i}\quad\quad\quad\quad\quad\quad\quad(i=0,...,K-1) \nonumber\\
Y_{i}&{}={}&G_{i}^{(1)}X_{1i}+G_{i}^{(3)}X_{2i}+Z_{i}\label{eq.4}
\end{eqnarray}
\setlength{\arraycolsep}{5pt}
where $Z_{i}\thicksim \mathcal{C}\mathcal{N}(0,N)$ and $Z_{1i}\thicksim \mathcal{C}\mathcal{N}(0,N_{1})$ are complex circularly symmetric zero mean additive white Gaussian noises with correlation coefficient $\rho_{z_{i}}{}={}\mathbb{E}\left\lbrace Z_{1i}Z_{i}^{*}\right\rbrace /\sqrt{NN_{1}}$.
\section{Upper and Lower Bounds on Capacity}
In this section, we provide the PDF achievable rate and max-flow min-cut upper bound for the capacity of the UWB relay channel.
\subsection{UWB Lower Bound}
When the channel between the source and the relay is better than the channel between the relay and the receiver, DF strategy gives the best achievable rate. To date the best rate achieved by DF strategy is obtained in \cite[theorem 7]{covergamal1979} by substituting  $\widehat{Y}_{1}{}={}\phi, V{}={} X_{2},U{}={}(X_{2}, U)$ which we call it as PDF. A delay-constrained form for this lower bound can be expressed as
\begin{small}
\setlength{\arraycolsep}{0.0em}
\begin{eqnarray}
C{}&\quad \geq& {}\sup _{(\textbf{X}_{\textbf{1}},\textbf{X}_{\textbf{2}},\textbf{U})}\min\bigg\lbrace \dfrac{1}{K} \sum_{i=0}^{K-1} I(X_{1i},X_{2i}; Y_{i}),\nonumber\\
&&\dfrac {1}{K} \sum_{i=0}^{K-1} I(U_{i};Y_{1i}\mid X_{2i}){}+{}I(X_{1i};Y_{i}\mid X_{2i}, U_{i})\bigg\rbrace \label{eq.7}
\end{eqnarray}
\setlength{\arraycolsep}{5pt}
\end{small}
where the supremum is taken over all joint probability mass functions of the form
\begin{equation}
\label{eq6.prob.distr} p(\textbf{X}_{\textbf{1}},\textbf{X}_{\textbf{2}},\textbf{U}){}={}\prod _{i=1}^{K}p(X_{2i})p(U_{i}\mid X_{2i})p(X_{1i}\mid U_{i}X_{2i})
\end{equation}\\
The resulted UWB version is expressed in the following theorem.
\paragraph*{Theorem 1}
A delay-constrained achievable rate with PDF strategy for frequency-selective block fading UWB relay channel is given by
\begin{equation}
R=\max _{
\setlength{\arraycolsep}{0.0em}
\begin{small} \begin{matrix} \overline{\alpha}_{0},&\dots ,&\overline{\alpha}_{k-1}\\
\overline{\beta}_{0},&\dots ,&\overline{\beta}_{k-1}\end{matrix}\end{small}
\setlength{\arraycolsep}{5pt}
}\min \left\lbrace \dfrac{1}{K}\sum_{i=0}^{K-1}C(\gamma_{1i}), \dfrac{1}{K}\sum_{i=0}^{K-1}C(\gamma_{2i})\right\rbrace ,\label{achievable}
\end{equation}
\begin{small}
\setlength{\arraycolsep}{0.0em}
\begin{eqnarray}
&\gamma_{1i}&{}={}\label{achievable1}\\
&&\dfrac{\vert G_{i}^{(1)}\vert^{2}P_{1}{}+{}\vert G_{i}^{(3)}\vert^{2} P_{2}{}+{}2\sqrt{P_{1}P_{2}}\Re \left\lbrace  \sqrt{\overline{\alpha}_{i}\overline{\beta}_{i}}G_{i}^{(1)}G_{i}^{(3)^{*}}\right\rbrace}
{N}\nonumber
\end{eqnarray}
\setlength{\arraycolsep}{5pt}
\end{small}
\begin{small}
\setlength{\arraycolsep}{0.0em}
\begin{eqnarray}
&\gamma_{2i}&{}={}\label{achievable2}\\
&&\left( 1+\dfrac{\big\vert G_{i}^{(2)}\big\vert^{2}\vert \alpha_{i}\vert \vert\overline{\beta}_{i}\vert P_{1}}{\big\vert G_{i}^{(2)}\big\vert^{2}\vert \beta_{i}\vert P_{1}+N_{1}}\right) \left( 1+\dfrac{\big\vert G_{i}^{(1)}\big\vert^{2}\vert \beta_{i}\vert P_{1}}{N}\right) {}-{}1\nonumber
\end{eqnarray}
\setlength{\arraycolsep}{5pt}
\end{small}
\paragraph*{Proof}
We now construct the code book and find the achievable rate based on the code book definitions. We assume that the source and relay nodes transmit their signals per complex baseband sample with average powers $P_{1}$ and $P_{2}$, respectively. For every $i\in\lbrace0, 1, \ldots, K-1\rbrace$ define $\overline{\alpha}_{i}$ and $\overline{\beta}_{i}$ as complex variables with $0\leq\vert\overline{\alpha}_{i}\vert, \vert\overline{\beta}_{i}\vert\leq1$ and assume that $\vert\alpha_{i}\vert{}={}1{}-{}\vert\overline{\alpha}_{i}\vert$ and $\vert\beta_{i}\vert{}={}1{}-{}\vert\overline{\beta}_{i}\vert$. Let $\setlength{\arraycolsep}{0.0em}
X_{2i}\thicksim \mathcal{C}\mathcal{N}(0,P_{2}), N_{1i}\thicksim \mathcal{C}\mathcal{N}(0,\vert\alpha_{i}\vert P_{0}), U_{i}\thicksim \mathcal{C}\mathcal{N}(0,P_{0}), M_{1i}\thicksim \mathcal{C}\mathcal{N}(0,\vert\beta_{i}\vert P_{1})\setlength{\arraycolsep}{5pt}
$ and $\setlength{\arraycolsep}{0.0em}
X_{1i}\thicksim \mathcal{C}\mathcal{N}(0,P_{1})\setlength{\arraycolsep}{5pt}
$, where $X_{2i},N_{1i}$ and $M_{1i}$ are mutually independent. First generate normal distributed random variables $X_{2i},N_{1i}$ and $M_{1i}$. Then, define $U_{i}$ and $X_{1i}$ as (\ref{eq6.prob.distr}) suggests in the following way
\begin{eqnarray}
U_{i}&{}={}&\sqrt{\overline{\alpha}_{i}\dfrac{P_{0}}{P_{2}}}X_{2i}{}+{}N_{1i}\\
X_{1i}&{}={}&\sqrt{\overline{\beta}_{i}\dfrac{P_{1}}{P_{0}}}U_{i}+M_{1i}\\
&{}={}&\sqrt{\overline{\alpha}_{i}\overline{\beta}_{i}\dfrac{P_{1}}{P_{2}}}X_{2i}{}+{}\sqrt{\overline{\beta}_{i}\dfrac{P_{1}}{P_{0}}}N_{1i}{}+{}M_{1i}\nonumber
\end{eqnarray}
The random code associated with this distribution is then given by
\setlength{\arraycolsep}{0.0em}
\begin{equation}
\begin{array}{@{}rcl@{\quad}rcl@{}}
\textbf{X}_{2}(s)i.i.d.&\thicksim & \mathcal{C}\mathcal{N}_{K}(0,P_{2}\textbf{I}) & s&\in &[1,2^{R_{0}}]\\
 \textbf{N}_{1}(w_{1})i.i.d.&\thicksim & \mathcal{C}\mathcal{N}_{K}(0,\textbf{C}_{\textbf{N}_{1}}) & w_{1}&\in &[1,2^{nR_{1}}]\\
 \textbf{M}_{1}(w_{2})i.i.d.&\thicksim & \mathcal{C}\mathcal{N}_{K}(0,\textbf{C}_{\textbf{M}_{1}}) & w_{2}&\in &[1,2^{nR_{2}}]
\end{array}
\end{equation}
\setlength{\arraycolsep}{5pt}
where the covariance matrices of $K$-variate complex normal distribution of $\textbf{N}_{\textbf{1}}$ and $\textbf{M}_{\textbf{1}}$ are
\begin{eqnarray}
&\textbf{C}_{\textbf{N}_{1}}{}={} diag(P_{0} \vert\alpha_{0}\vert,P_{0}  \vert\alpha_{1}\vert, \cdots, P_{0} \vert \alpha_{K-1}\vert )&\\
&\textbf{C}_{\textbf{M}_{1}}{}={}diag(P_{1} \vert\beta_{0}\vert, P_{1} \vert\beta_{1}\vert, \cdots, P_{1}\vert \beta_{K-1}\vert ),&
\end{eqnarray}
and $\textbf{U}$ and $\textbf{X}_{1}$ are constructed as
\begin{small}
\setlength{\arraycolsep}{0.0em}
\begin{eqnarray}
\textbf{U}(w_{1}\mid s)&=&\sqrt{\dfrac{P_{0}}{P_{2}}}[\overline{\alpha}_{0} \;  \overline{\alpha}_{1} \; \cdots \; \overline{\alpha}_{K-1}]\times \textbf{X}_{2}(s)\\
&&{}+{}\textbf{N}_{1}(w_{1})\nonumber\\
\textbf{X}_{1}(w_{2}\mid w_{1},s)&{}={}&\sqrt{\dfrac{P_{1}}{P_{2}}}[\overline{\alpha}_{0} \overline{\beta}_{0}\;  \overline{\alpha}_{1}\overline{\beta}_{1}\;  \cdots \;\overline{\alpha}_{K-1}\overline{\beta}_{K-1}]\times \textbf{X}_{2}(s)\nonumber\\
&&{}+{}\sqrt{\dfrac{P_{1}}{P_{0}}}[\overline{\beta}_{0}\; \overline{\beta}_{1}\;  \cdots \; \overline{\beta}_{K-1}]\times\textbf{N}_{1}(w_{1})\nonumber\\
&&{}+{}\textbf{M}_{1}(w_{2})
\end{eqnarray}
\setlength{\arraycolsep}{5pt}
\end{small}
where $\times$ denotes an element by element matrix multiplication.
Then if
\begin{small}
\setlength{\arraycolsep}{0.0em}
\begin{eqnarray}
R_{0}&<&\frac{1}{K}\sum_{i=0}^{K} I(X_{2i},Y_{i})\\
R_{1}&<&\frac{1}{K}\sum_{i=0}^{K}\min\left\lbrace I\left( U_{i};Y_{1i}|X_{2i}\right),R_{0}+I\left(U_{i};Y_{i}|X_{2i}\right) \right\rbrace \\
R_{2}&<&\frac{1}{K}\sum_{i=0}^{K-1}I\left(X_{1i};Y_{i}|U_{i},X_{2i}\right)
\end{eqnarray}
\setlength{\arraycolsep}{5pt}
\end{small}
where
\begin{small}
\setlength{\arraycolsep}{0.0em}
\begin{eqnarray}
\label{eq.16}I(&X_{2i}&;Y_{i}){}={}h(Y_{i})-h(Y_{i}\mid X_{2i})\nonumber\\
&{}={}&\log \big(\pi e\; var(Y_{i})\big){}-{}\log \big(\pi e\; \mathbb{E}\;var(Y_{i}\mid X_{2i})\big) \nonumber\\
&{}={}&\log\left( 1+\frac{\left|G_{i}^{(1)}\sqrt{\overline{\alpha}_{i}\overline{\beta}_{i}P_{1}}{}+{}G_{i}^{(3)}\sqrt{P_{2}}\right|^{2}}{\left|G_{i}^{(1)}\right|^{2}P_{1}\left(\left|\alpha_{i}\right| \left|\overline{\beta}_{i}\right|+\left|\beta_{i}\right|\right)+N}\right),\\
I(&U_{i}&;Y_{1i}\mid X_{2i}){}={}h(Y_{1i}\mid X_{2i})-h(Y_{1i}\mid X_{2i},U_{i})\nonumber\\
&{}={}&\log \bigg(\pi e\; \mathbb{E}\;var(Y_{1i}\mid X_{2i})\bigg)\nonumber\\
&&{}-{}\log \bigg(\pi e\;\mathbb{E}\; var(Y_{1i}\mid X_{2i},U_{i})\bigg)\nonumber\\
&{}={}&\log \bigg(1+\dfrac{\big\vert G_{i}^{(2)}\big\vert^{2}\vert\alpha_{i}\vert \vert\overline{\beta}_{i}\vert P_{1}}{\big\vert G_{i}^{(2)}\big\vert^{2}\vert\beta_{i}\vert P_{1}{}+{}N_{1}}\bigg),\\
I(&U_{i}&;Y_{i}\mid X_{2i}){}={}h(Y_{i}\mid X_{2i})-h(Y_{i}\mid X_{2i},U_{i})\nonumber\\
&{}={}&\log \bigg(\pi e\mathbb{E}var(Y_{i}\mid X_{2i})\bigg){}-{}\log \bigg(\pi e\mathbb{E} var(Y_{i}\mid X_{2i},U_{i})\bigg)\nonumber\\
&{}={}&\log \bigg(1+\dfrac{\big\vert G_{i}^{(1)}\big\vert^{2}\vert\alpha_{i}\vert \vert\overline{\beta}_{i}\vert P_{1}}{\big\vert G_{i}^{(1)}\big\vert^{2}\vert\beta_{i}\vert P_{1}{}+{}N}\bigg),\\
I&(X_{1i}&;Y_{i}\mid X_{2i},U_{i}){}={}h(Y_{i}\mid X_{2i},U_{i})-h(Y_{i}\mid X_{1i},X_{2i},U_{i})\nonumber\\
&{}={}&\log \bigg(\pi e\; \mathbb{E}\;var(Y_{i}\mid X_{2i},U_{i})\bigg)\nonumber\\
&&{}-{}\log \bigg(\pi e\; \mathbb{E}\;var(Y_{i}\mid X_{1i},X_{2i},U_{i})\bigg)\nonumber\\
&{}={}&\log \bigg(1+\dfrac{\big\vert G_{i}^{(1)}\big\vert^{2}\vert\beta_{i}\vert P_{1}}{N}\bigg),\label{eq.17} 
\end{eqnarray}
\setlength{\arraycolsep}{5pt}
\end{small}
the rate $R=R_{1}+R_{2}$ can be achieved with arbitrarily small probability of error and the proof of Theorem 1 is completed.
\subsection{UWB Upper Bound}
A $K$-block delay constrained form for the max-flow min-cut upper bound on the capacity of the general relay channel, established in \cite[theorem 7]{covergamal1979} can be expressed as
\setlength{\arraycolsep}{0.0em}
\begin{eqnarray}
\label{upperbound}C\leq \sup_{p(X_{1},X_{2})}\min\bigg\lbrace\dfrac{1}{K}&\sum_{i=0}^{K-1}&I(X_{1i},X_{2i};Y_{i}),\\
\dfrac{1}{K}&\sum_{i=0}^{K-1}&I(X_{1i};Y_{i},Y_{1i}\mid X_{2i})\bigg\rbrace\nonumber
\setlength{\arraycolsep}{5pt}\end{eqnarray}
The UWB version of this upper bound is expressed in the following theorem.

\paragraph*{Theorem 2} 
The delay-constrained max-flow min-cut upper bound on the capacity of a frequency-selective block fading relay channel is given by
\begin{small}
\begin{eqnarray}
C{}\leq{}\max_{\setlength{\arraycolsep}{0.0em}\begin{small}\begin{matrix} \overline{\alpha}_{0},&\dots,&\overline{\alpha}_{k-1}\\
\overline{\beta}_{0},&\dots, &\overline{\beta}_{k-1}\end{matrix}\end{small}}\min\left\lbrace \frac{1}{K}\sum_{i=0}^{K-1}C(\gamma_{1i}), \frac{1}{K}\sum_{i=0}^{K-1}C(\gamma_{3i})\right\rbrace, \label{eq.uwb upper bound}
\end{eqnarray}
\end{small}
where
\setlength{\arraycolsep}{0.0em}
\begin{eqnarray}
\gamma_{3i}{}={}&&P_{1}\dfrac{(1-\vert\overline{\alpha}_{i}\vert\vert\overline{\beta}_{i}\vert)}{1-\vert\rho_{zi}\vert^{2}}\\
&\bigg(&\dfrac{\vert G_{i}^{(1)}\vert^{2}}{N}+\dfrac{\vert G_{i}^{(2)}\vert^{2}}{N_{1}}-2\dfrac{\Re\big\lbrace G_{i}^{(1)}G_{i}^{(2)^{*}}\rho_{zi}\big\rbrace}{\sqrt{NN_{1}}}\bigg)\nonumber
\end{eqnarray}
\setlength{\arraycolsep}{5pt}
and $\gamma_{1i}$ has been defined in (\ref{achievable1}).
\paragraph*{Proof}
We start from (\ref{upperbound}). By noticing that normal random variables have the maximum entropy and by letting $X_{1i}$ and $X_{2i}$ each have the maximum allowed power $P_{1}$ and $P_{2}$ and by choosing $E(X_{1i}X_{2i}^{*})=\sqrt{\overline{\alpha}_{i}\overline{\beta}_{i}P_{1}P_{2}}$, the first term $I(X_{1i},X_{2i};Y_{i})$ is upper bounded by $C(\gamma_{1i})$, the same expression as in the lower bound. The proof is a trivial extension of that in \cite[theorem 4]{covergamal1979} and is skipped here. For the second term, by the same assumptions we have
\setlength{\arraycolsep}{0.0em}
\begin{eqnarray}
I(&X_{1i}&;Y_{i},Y_{1i}\mid X_{2i}){}={}h(Y_{i},Y_{1i}\mid X_{2i})-h(Z_{i},Z_{1i})\nonumber\\
&\leq &\dfrac{1}{2}\log \big( (2\pi e)^{2}\mathbb{E}\;\det cov(Y_{i},Y_{1i}\mid X_{2i})\big)\nonumber\\
&&-\dfrac{1}{2}\log\big((2\pi e)^{2}\mathbb{E}\;\det cov(Z_{i},Z_{1i} \big)=C(\gamma_{3i})
\end{eqnarray}
where
\begin{eqnarray}
\mathbb{E}\; &\det\; &cov(Y_{i},Y_{1i}\mid X_{2i})\nonumber\\
&=&NN_{1}\big(1-\vert\rho_{z_{i}}\vert^{2}\big)+P_{1}NN_{1}\left(1-\vert\overline{\alpha}_{i}\vert \vert\overline{\beta}_{i}\vert\right)\nonumber\\
&&\bigg(\frac{\big\vert G_{i}^{(1)}\big\vert^{2}}{N}+\frac{\big\vert G_{i}^{(2)}\big\vert ^{2}}{N_{1}}-2\frac{\;\Re\left\lbrace G_{i}^{(1)}G_{i}^{(2)^{*}}\rho_{z_{i}}\right\rbrace}{\sqrt{NN_{1}}}\bigg),\nonumber\\
\mathbb{E}\; &\det\; & cov(Z_{i},Z_{1i} \big)=NN_{1}\left( 1-\vert\rho_{z_{i}}\vert ^{2}\right)
\end{eqnarray}\setlength{\arraycolsep}{5pt}
Due to space constraints, the details are omitted.
\setlength{\arraycolsep}{0.0em}
\section{Capacity of Degraded UWB Relay Channel}
\paragraph*{Theorem 3}
The delay-constrained capacity of the degraded frequency-selective block fading relay channel is
\setlength{\arraycolsep}{0.0em}
\begin{eqnarray}
C{}={}\max_{\overline{\alpha}_{0},\ldots, \overline{\alpha}_{k-1}}\min\bigg\lbrace\dfrac{1}{K}\sum_{i=0}^{K-1}\log\bigg(1+\frac{1}{N}\big(\vert G_{i}^{(1)}\vert^{2}&P_{1}&\nonumber\\
+\vert G_{i}^{(3)}\vert^{2}P_{2}+2\sqrt{P_{1}P_{2}}\;\Re \bigg \lbrace\sqrt{\overline{\alpha}_{i}}G_{i}^{(1)}G_{i}^{(3)^{*}}\bigg \rbrace\big)&\bigg)&\nonumber\\
,\dfrac{1}{K}\sum_{i=0}^{K-1}\log\bigg(1+\dfrac{\vert G_{i}^{(2)}\vert^{2}\vert\alpha_{i}\vert P_{1}}{N_{1}}\bigg)&\bigg\rbrace &
\end{eqnarray}
\setlength{\arraycolsep}{5pt}
\paragraph*{Proof}
We evaluate the upper and lower bounds on the capacity of the degraded UWB relay channel and show that they coincide with each other.
As stated in \cite{covergamal1979} a relay channel is called degraded if the following relationship holds
\begin{equation}
p(y\mid y_{1},x_{1},x_{2}){}={}p(y\mid y_{1},x_{2})\label{eq.degraded}
\end{equation}

To adapt the general input-output relation (\ref{eq.4}) to this condition, we rewrite (\ref{eq.4}) as
\setlength{\arraycolsep}{0.0em}
\begin{eqnarray}
Y_{1i}&{}={}&G_{i}^{(2)}X_{1i}+Z_{1i}\nonumber\\
Y_{i}&{}={}&\dfrac{G_{i}^{(1)}}{G_{i}^{(2)}}Y_{1i}{}+{}G_{i}^{(3)}X_{2i}{}+{}Z_{2i}\end{eqnarray}
\setlength{\arraycolsep}{0pt}
where
\begin{equation}
Z_{2i}{}={}Z_{i}-\dfrac{G_{i}^{(1)}}{G_{i}^{(2)}}Z_{1i}
\end{equation}

In order to hold (\ref{eq.degraded}), $Z_{1i}$ and $Z_{2i}$ should be independent for each $i$. For normal distributed random variables correlation and dependency are equivalent terms, thus we find $\rho_{zi}$ so that $Z_{1i}$ and $Z_{2i}$ become uncorrelated. This is achieved when\\
\begin{equation}
\rho_{zi}{}={}\left( \frac{G_{i}^{(1)}}{G_{i}^{(2)}}\right) ^{*}\sqrt{\frac{N_{1}}{N}}\label{eq.rodeg}
\end{equation}
\paragraph*{Achievability}
The achievable rate is resulted from the substitution of $U_{i}{}={}\sqrt{\frac{P_{0}}{P_{1}}}X_{1i}$ in the code book of Theorem 1, i.e., by setting $\overline{\beta}_{i}{}={}1$ in (\ref{achievable}),  (\ref{achievable1}) and  (\ref{achievable2}). 
\paragraph*{Converse}
We have computed the max-flow min-cut upper bound in the previous section for the general UWB relay channel. The upper bound on the capacity of degraded UWB relay channel can be obtained if we apply the condition (\ref{eq.rodeg}) to the established upper bound (\ref{eq.uwb upper bound}) and to make the notations consistent with the achievability part, choosing $E(X_{1i}X_{2i}^{*})=\sqrt{\overline{\alpha}_{i}P_{1}P_{2}}$ in the proof of the upper bound. By applying these substitutions the obtained upper bound coincides with the lower bound and the proof is completed.

\section{Capacity of Reversely Degraded UWB Relay Channel}
\paragraph*{Theorem 4}
The delay-constrained capacity of the reversely degraded frequency-selective block fading relay channel is
\setlength{\arraycolsep}{0.0em}
\begin{eqnarray}
C{}={}\dfrac{1}{K}\sum_{i=0}^{K-1}\log\bigg(1+\dfrac{\vert G_{i}^{(1)}\vert^{2} P_{1}}{N}\bigg)  
\end{eqnarray}
\setlength{\arraycolsep}{5pt}
\paragraph*{Proof}
As stated in \cite{covergamal1979} a relay channel is called reversely degraded if the following relationship holds
\begin{equation}
p(y_{1}\mid x_{1},x_{2},y){}={}p(y_{1}\mid y,x_{2})
\label{revdegraded} \end{equation}
To adapt the general input-output relation (\ref{eq.4}) to this constraint, we rewrite (\ref{eq.4}) as
\begin{eqnarray}
Y_{1i}&{}={}&\dfrac{G_{i}^{(2)}}{G_{i}^{(1)}}Y_{i}{}-{}\dfrac{G_{i}^{(2)}G_{i}^{(3)}}{G_{i}^{(1)}}X_{2i}{}+{}Z_{3i}\\
Y_{i}&{}={}&G_{i}^{(1)}X_{1i}{}+{}G_{i}^{(3)}X_{2i}+Z_{i},\quad \nonumber
\end{eqnarray}
where
\begin{equation}
Z_{3i}{}={}Z_{1i}-\dfrac{G_{i}^{(2)}}{G_{i}^{(1)}}Z_{i}
\end{equation}
In order to hold (\ref{revdegraded}), $Z_{i}$ and $Z_{3i}$ should be independent for each $i$. This is achieved if\\
\begin{equation}
\rho_{zi}{}={}\frac{G_{i}^{(2)}}{G_{i}^{(1)}}\sqrt{\frac{N}{N_{1}}} \label{eq.rorevdeg}
\end{equation}
\paragraph*{Achievability}
The achievable rate is resulted from the substitution of $U_{i}{}={}\sqrt{\frac{P_{0}}{P_{2}}}X_{2i}$ in the code book of Theorem 1, i.e., by substituting $\overline{\alpha}_{i}{}={}1$ in (\ref{achievable}),  (\ref{achievable1}) and (\ref{achievable2}).

\setlength{\arraycolsep}{5pt}
\paragraph*{Converse}
The upper bound on the capacity of reversely degraded UWB relay channel can be obtained if we apply the condition (\ref{eq.rorevdeg}) to the established upper bound (\ref{eq.uwb upper bound}) and to make the notations consistent, choosing $E(X_{1i}X_{2i}^{*})=\sqrt{\overline{\beta}_{i}P_{1}P_{2}}$. By applying these substitutions the upper bound is reached and coincides with the lower bound as follows
\setlength{\arraycolsep}{0.0em}
\begin{small}
\begin{eqnarray}
\label{eq.revdeg previous} C{}={}\sup_{\overline{\beta}_{0},\ldots, \overline{\beta}_{k-1}}\min\bigg\lbrace\dfrac{1}{K}\sum_{i=0}^{K-1}\log\bigg(1+\frac{1}{N}\big(\vert G_{i}^{(1)}\vert^{2}&P_{1}&\nonumber\\
+\vert G_{i}^{(3)}\vert^{2}P_{2}+2\sqrt{P_{1}P_{2}}\;\Re \bigg \lbrace\sqrt{\overline{\beta}_{i}}G_{i}^{(1)}G_{i}^{(3)^{*}}\bigg \rbrace \big)&\bigg)&\nonumber\\
,\dfrac{1}{K}\sum_{i=0}^{K-1}\log\bigg(1+\dfrac{\vert G_{i}^{(1)}\vert^{2}\vert\beta_{i}\vert P_{1}}{N}\bigg)&\bigg\rbrace & 
\end{eqnarray}
\end{small}
Now, we show that the first term in (\ref{eq.revdeg previous}) is always greater than the second one. Consider $S$ as the subtraction of the second term from the first term of (\ref{eq.revdeg previous}), determined as
\begin{equation}
S=\dfrac{1}{K}\sum_{i=0}^{K-1}C(\zeta_{i})
\end{equation} 
where
\begin{small}
\setlength{\arraycolsep}{0.0em}
\begin{eqnarray}
\zeta_{i}=&&\\
&&\dfrac{\big\vert G_{i}^{(1)}\big\vert^{2}\vert \overline{\beta}_{i}\vert P_{1}+\big\vert G_{i}^{(3)}\big\vert^{2}P_{2}+2\sqrt{P_{1}P_{2}}\;\Re\left\lbrace \sqrt{\overline{\beta}_{i}}G_{i}^{(1)}G_{i}^{(3)^{*}}\right\rbrace}{N+K\big\vert G_{i}^{(1)}\big\vert^{2}\vert\beta_{i}\vert P_{1}}\nonumber
\end{eqnarray}
\setlength{\arraycolsep}{5pt}
\end{small}
We now show that $\zeta_{i}\geq 0$. The minimum value of $\zeta_{i}$ happens when $\sqrt{\overline{\beta}_{i}}G_{i}^{(1)}G_{i}^{(3)^{*}}$ is a real negative value, so considering the worst case, we choose
\begin{equation}
\sqrt{\overline{\beta}_{i}}=-\big\vert\sqrt{\overline{\beta}_{i}}\big\vert\dfrac{G_{i}^{(1)^{*}}G_{i}^{(3)}}{\vert G_{i}^{(1)^{*}}G_{i}^{(3)}\vert}.
\end{equation}
Now, the numerator of $\zeta_{i}$ can be written as
\begin{equation}
\left(\big\vert G_{i}^{(1)}\big\vert\big\vert\sqrt{\overline{\beta}_{i}}\big\vert\sqrt{P_{1}}-\big\vert G_{i}^{(3)}\big\vert\sqrt{P_{2}} \right)^{2}\geq0  
\end{equation}
and the denominator of $\zeta_{i}$ is always positive. Therefore, $\zeta_{i}\geq0$ and $S\geq0$. So, the minimum of the two terms of (\ref{eq.revdeg previous}) is the second one and the capacity is achieved by maximizing the second term with respect to $\overline{\beta}_{i}$ which results in $\overline{\beta}_{i}=0$. This completes the proof of the Theorem 4.
\section{Numerical Results}
\begin{figure}[!t]
\centering
\includegraphics[width=3.2in]{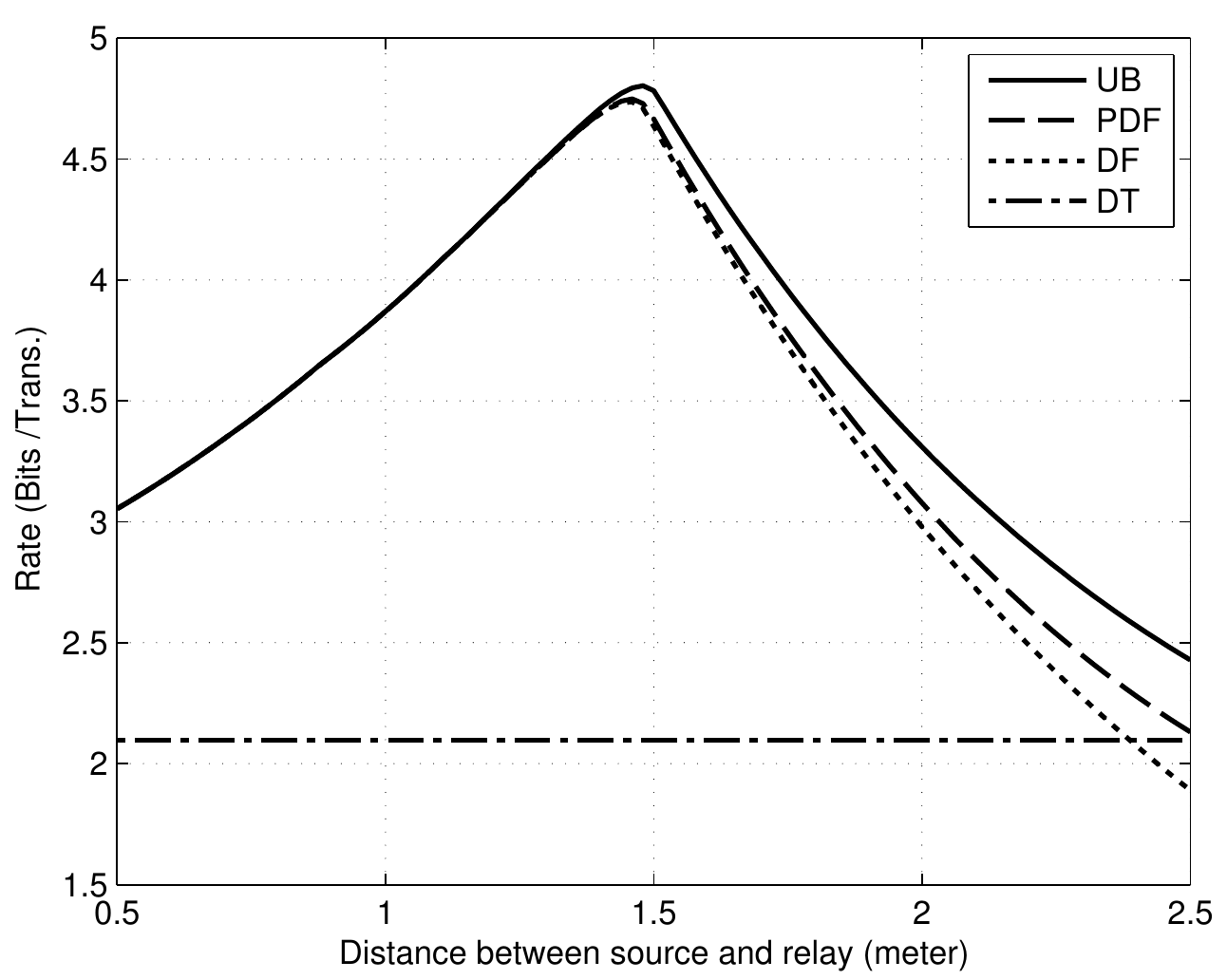}
\caption{Comparisons of the bounds on the UWB relay channel capacity}
\label{figure2}
\end{figure}
\begin{figure}[!t]
\centering
\includegraphics[width=3.2in]{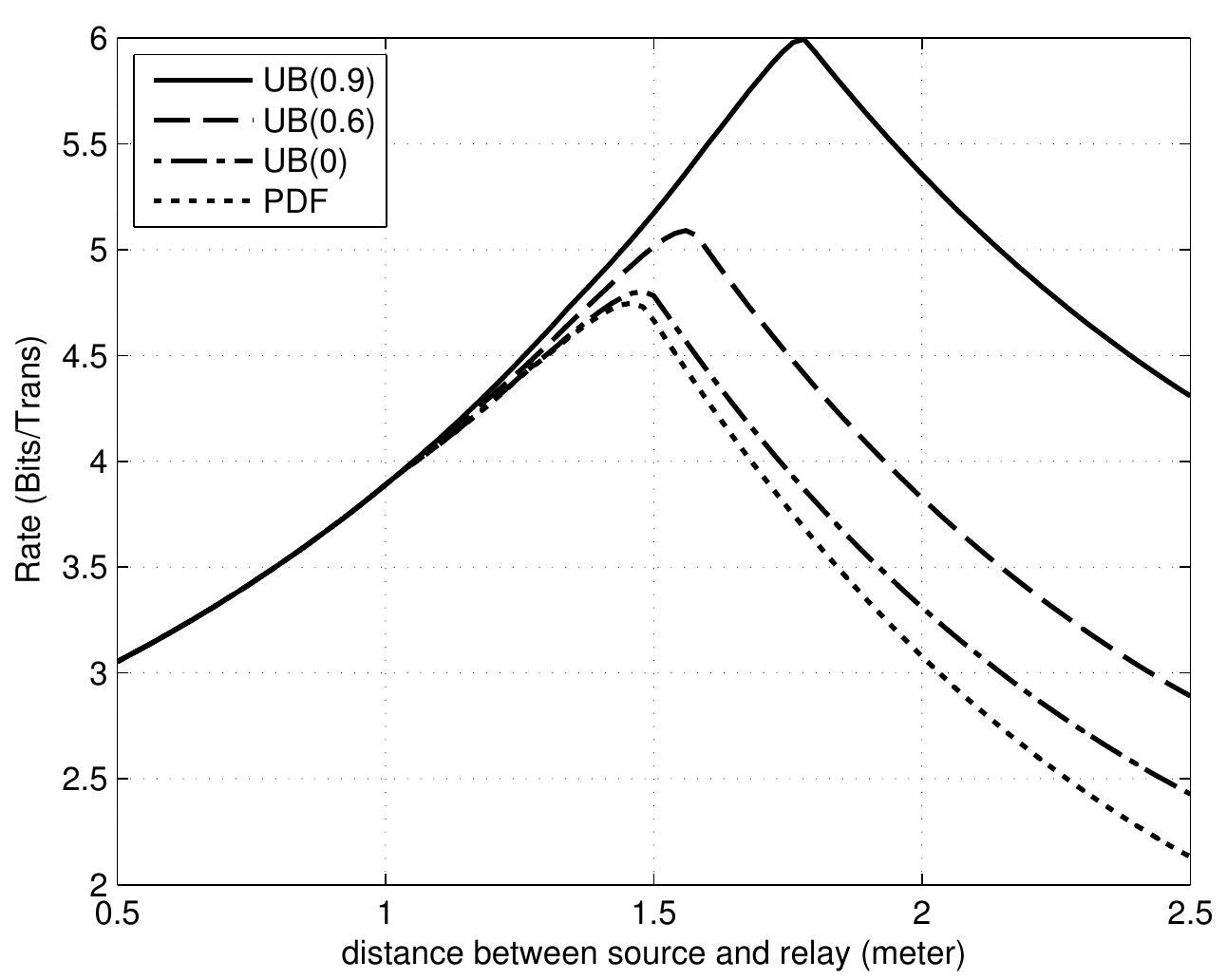}
\caption{PDF and Upper bound for different values of $\rho_{zi}$}
\label{figure3}
\end{figure}
In this section, we present illustrating figures to examine the obtained general achievable rate (PDF) and the max-flow min-cut upper bound and compare them with the DF achievable rate which is a special form of PDF by substituting $\overline{\beta}_{i}=0$ which was investigated in \cite{zolfa2009}. The simulations are done based on the channel model for residential NLOS environments in \cite{ieee802.15.4a}. The transmitted powers by the source and the relay nodes are equal to the maximum allowed power for the UWB systems, defined by FCC ($-$41.3dBm/MHz). We assume equal noise power spectral densities at the relay and destination($-$114dBm/MHz). The distance between the source and destination is fixed at $d_{1}=3m$ and the bounds are plotted versus the distance between the source and the relay. In Fig. \ref{figure2}, we set $\rho_{zi}=0$ in the upper bound. As we see when the multiple-access channel is the bottleneck, the three bounds reach the same rate. The difference occurs when the broadcast channel is the bottleneck in which PDF performs better than the DF and this improvement increases as the relay moves toward the destination. The capacity of the direct transmission is also plotted in which the power of the source is assumed twice its power in the relay channel scenario for a more fair comparison. In Fig. \ref{figure3} the upper bound for three different values of $\rho_{zi}=0, 0.6, 0.9$ is plotted. It can be observed that the upper bound increases for higher values of correlation coefficients and also the peak value occurrence of the upper bound moves toward the destination.
\section*{Conclusion}
In this paper, we first computed an achievable rate obtained with PDF coding scheme and the max-flow min-cut upper bound with correlated noises at the relay and destination for a relay channel with UWB links versus channel coefficients, transmitted powers and correlation coefficients of noises with the assumption of known CSI at the receiving terminals only. Then, by appropriately finding the corresponding noise correlation coefficients, we established the capacity of degraded and reversely degraded UWB relay channels. The UWB lower and upper bounds obtained here can be used for further investigation of the UWB relay channels, specially for the ones with known capacities. 

\bibliographystyle{IEEEtran}	
\bibliography{citation1}

\end{document}